\documentstyle[11pt]{article}

\def\ar{\rightarrow}
\def\bga{B_{\al}\,^{\ga}}

\def\bbga{B_{\al}\,^{\ga\de}}
\def\bbep{B_{\al}\,^{\ep\ze}}
\def\bib{\bibitem}
\def\ccga{C_{\al}\,^{\ga\de}}

\def\ega{e_{\al}\,^{\ga}}

\def\emu{e_{\al}\,^{\mu}}

\def\dga{\delta_{\al}\,^{\ga}}
\def\dem{\det e^{-1}}

\def\hga{H_{\al\be}\,^{\ga}}

\def\gmu{g^{\mu\nu}}
\def\intf{\int d^{4}x\,}

\def\lar{\longrightarrow}
\def\lbr{\lbrack}

\def\mga{m_{\ga\de}}
\def\pa{\partial}

\def\pga{p_{\ga}}
\def\rbr{\rbrack}
\def\psiba{\overline{\psi}}
\def\tr{\,\mbox{tr}\,}
\def\Tr{\,\mbox{Tr}\,}
\def\al{\alpha}
\def\be{\beta}
\def\ga{\gamma}
\def\de{\delta}
\def\ep{\varepsilon}
\def\ze{\zeta}
\def\th{\vartheta}

\def\la{\lambda}
\def\La{{\it\Lambda}}
\def\Ph{{\it\Phi}}

\def\Si{\Sigma}
\def\va{\varphi}
\def\om{\omega}

\def\beq{\begin{equation}}
\def\eeq{\end{equation}}
\def\bed{\begin{displaymath}}
\def\eed{\end{displaymath}}
\def\beqq{\begin{eqnarray}}
\def\eeqq{\end{eqnarray}}
\def\bedd{\begin{eqnarray*}}
\def\eedd{\end{eqnarray*}}

\title{\huge{\bf{Poincar$\acute{\mbox{e}}$ gauge invariance \\ and \\
gravitation in Minkowski spacetime}}}
\author{\vspace{0.5cm}\\
C. Wiesendanger\thanks{e-mail address: wie@stp.dias.ie}\\
Dublin Institute for Advanced Studies\\
School of Theoretical Physics\\
10 Burlington Road\\
Dublin 4, Ireland}
\date{May 5, 1995}

\begin{document}

\maketitle

\begin{abstract}
A formulation of Poincar$\acute{\mbox{e}}$ symmetry as an inner symmetry
of field theories defined on a fixed Minkowski spacetime is given. Local {\bf
P\/}
gauge transformations and the corresponding covariant derivative with {\bf P\/}
gauge fields are introduced. The renormalization properties of scalar, spinor
and
vector fields in {\bf P\/} gauge field backgrounds are determined. A minimal
gauge
field dynamics consistent with the renormalization constraints is given.\\

\end{abstract}

\vspace{1cm}

\begin{flushleft}
{\large DIAS-STP-95-18}
\end{flushleft}

\clearpage

\section{Introduction}

\paragraph{}
In physics the description of a class of phenomena may often be based on
different a
priori conventions, hence yielding complementary pictures of these phenomena.
In this
context Poincar$\acute{\mbox{e}}$ pointed out the purely conventional role of
spacetime
geometry in the description of the behaviour of matter \cite{poi}. In fact two
points
of view are possible \cite{mit}.

Either, one defines the line element $ds^{2}$ to be of Minkowskian form.
Accordingly,
in a gravitational field material rods will shrink and clocks slow down w.r.t.
this
metric. Hence, one defines the geometry of spacetime to be Minkowskian, whereas
the
behaviour of physical rods and clocks has to be determined by experiments.

Or, one defines rods or clocks to have one and the same length or period at any
point of spacetime. Accordingly, a measurement of the line element $ds^{2}$
using
these rods and clocks will yield that the geometry of spacetime is curved in
general.
This is the convention Einstein introduced to describe gravitation. Apart from
global
topological questions the two complementary points of view are equivalent.

The general theory of relativity and its extensions are based on the second
point of
view and yield a geometric description of the gravitational interaction
\cite{sexu}.
This is incorporated in the theory by requiring that the behaviour of matter in
gravitational backgrounds has to be described by equations which are
form-invariant
under the groups of general coordinate transformations and local $SO(1,3)$
frame
rotations \cite{sexu}. In this conception the gravitational field is closely
connected
to the metric tensor.

As long as one is interested in the macroscopic aspects of gravitation this
point of
view is very natural \cite{str}. Its limitation shows up at the quantum level.
It is very
difficult to extend a picture so intimately related to classical concepts such
as
rods and clocks to a simple microscopic understanding of gravitation. In
microphysics
spacetime geometry enters only as a background concept necessary in defining a
field theory. It cannot be subject to direct measurements in this context.

Hence, at the quantum level one is naturally led to the first point of view
avoiding
the interrelation of spacetime structure and gravitational phenomena. Here free
matter
is described by local, causal fields defined on Minkowski spacetime and its
interactions are introduced using the gauge principle which allows a
far-reaching
generalization of the connection between conservation laws and global symmetry
requirements \cite{lor}.

To obtain a gauge theory of gravitation \cite{iva} one first ensures the
conservation of
energy-momentum and angular momentum by the requirement of global covariance of
the
free matter field theory under the Poincar$\acute{\mbox{e}}$ group {\bf P\/}.
In a
second step one gauges {\bf P\/} \cite{kib}-\cite{kik2}, the translation
subgroup
{\bf T\/} \cite{hay1}-\cite{kik1}, the Lorentz subgroup {\bf L\/} \cite{uti},
or even
a larger group, e.g. \cite{yan}-\cite{lord}.

Most of the existing gauge theories of gravitation adopt
the second point of view yielding a geometric description of gravity
\cite{kib}-\cite{hen}, \cite{hay1}. This is related to
the fact that {\bf P\/} is usually conceived partly as a spacetime partly as an
inner symmetry group. The local extension of its spacetime part becomes then
the
diffeomorphism group, such that the gauged theory is invariant under general
coordinate transformations and local $SO(1,3)$ frame rotations. This local
symmetry
group is then necessarily linked with the geometry of spacetime.

Adopting the second point of view one can build up gauge theories of
gravitation
erecting a principal bundle with {\bf T\/} \cite{cho1}-\cite{kik1},
{\bf P\/} \cite{cho2}-\cite{kik2} or some other group as structural group
\cite{man}-\cite{lord}.
One difficulty is now to link the connection corresponding to the purely inner
symmetry with the vierbein and spin connection in gravity. Another one comes
with
choosing an action for the gauge fields natural from the bundle point of view.
It may come out to be inconsistent with renormalization properties of matter
fields
in such backgrounds.

Therefore we restrict ourselves to recast {\bf P\/} symmetry and its
consequences
in the form of an inner symmetry (section 2) extending a previous work on
gauging
the translations alone \cite{wie}. This leads to a complementary
description of the global action of {\bf P\/} which is in complete analogy to
the
description of the action of inner symmetry groups as groups of generalized
'rotations' in field space \cite{lor}. In particular the coordinate system used
to
specify the spacetime events is not affected anymore by {\bf P\/}
transformations.

We next introduce local  {\bf P\/} gauge transformations and demand the
invariance
of physical processes under those (sections 3 and 4). This necessarily leads to
the
existence of gauge fields with definite behaviour under local {\bf P\/} gauge
transformations. Their coupling to any other field is essentially fixed as in
the
case of other gauge field theories (section 5).

To obtain a gauge field dynamics consistent with renormalization properties of
matter fields we next determine the changes of one-loop partition functions
under rescaling (section 7). In a renormalizable theory the anomalous
contributions
to these changes may be absorbed in the classical actions for the different
fields
(e.g. \cite{wip}.
Using heat kernel methods and the $\zeta$-function renormalization for one-loop
determinants shortly presented in appendices A and B we determine the
contributions
of the two gauge fields explicitly. We finally give a minimal gauge field
action
consistent with theses contributions (section 8).

In order to get the interpretation of the resulting theory as a gauge theory of
gravitation we show that it may be recast in usual geometrical terms replacing
local {\bf P\/} gauge invariance by invariance under general coordinate
transformations and local $SO(1,3)$ frame rotations, the symmetry requirements
in the general theory of relativity or its extensions (section 6). Hence, the
{\bf P\/}
gauge theory of gravitation allows a complementary description of gravitational
effects in which the mathematical structure of the underlying spacetime is not
affected by physical events (in this context we refer to \cite{gri} -
\cite{pet}).

We work on Minkowski spacetime ({\bf R\/}$^{4}$,$\eta$) with Cartesian
coordinates
throughout, such that $\eta=\mbox{diag}(1,-1,-1,-1)$. Indices $\al,\be,\ga,...$
from the first half of the Greek alphabet denote quantities defined on
({\bf R\/}$^{4}$,$\eta$) which transform covariantly w.r.t. the Lorentz group.
They are correspondingly raised and lowered with $\eta$.

\section{Global Poincar$\acute{\mbox{e}}$ invariance as an inner symmetry}

\paragraph{}
In this section we extend the two complementary conceptions of global
translation
invariance in field theory discussed in \cite{wie} to the full
Poincar$\acute{\mbox{e}}$ group
including fields with spin. The corresponding conserved currents, the canonical
energy-momentum tensor $\Theta^{\ga}\,_{\al}$ and the angular momentum tensor
${\cal M}^{\ga}\,_{\al\be}$ coincide in the two conceptions.

First let us state the Noether theorem in a quite general form.
Consider a set of fields $\va_{j}(x)$ with $j=1,..,n$. Their dynamics shall be
specified by the action $S_{M}=\intf\,{\cal
L}_{M}(x,\va_{j},\pa_{\al}\va_{j})$.
$\de S_{M}=0$ yields then the equations of motion. Consider further the
infinitesimal
transformations
\beqq x^{\al}\lar x'^{\al}&=&x^{\al}+\de x^{\al}(x), \\
\va_{j}(x)\lar \va_{j}'(x')&=&\va_{j}(x)+\de\va_{j}(x) \nonumber \eeqq
of the coordinates and the fields. If there are functions $\de f^{\ga}(x)$ for
which
\beqq \label{2} & &\quad d^{4}x'\,{\cal
L}_{M}(x',\va_{j}'(x'),\pa'_{\al}\va_{j}'
(x'))= \\ & &d^{4}x\,\{{\cal L}_{M}(x,\va_{j}(x),\pa_{\al}\va_{j}(x))+
\pa_{\ga}\de f^{\ga}(x)\}\nonumber \eeqq
holds, then there is a conserved current $J^{\ga}$ found to be
\beq \label{3}J^{\ga}=-\frac{\pa{\cal
L}_{M}}{\pa(\pa_{\ga}\va_{j})}\cdot\de\va_{j}
+\de f^{\ga}+\Theta^{\ga}\,_{\al}\cdot\de x^{\al} \eeq
and, with appropriate boundary conditions, a conserved charge $Q$ given by
\beq Q=\int\limits_{x^{0}=const}^{} d^{3}x\, J^{0}.\eeq
The fields $\va_{j}$ above must obey the equations of motion.
$\Theta^{\ga}\,_{\al}$ is the canonical energy-momentum tensor
\beq \Theta^{\ga}\,_{\al}=\frac{\pa{\cal L}_{M}}{\pa(\pa_{\ga}\va_{j})}\cdot
\pa_{\al}\va_{j}-\eta^{\ga}\,_{\al}\cdot{\cal L}_{M}.\eeq
We apply this theorem now in two different ways to a globally
Poincar$\acute{\mbox{e}}$
invariant theory of the fields $\va_{j}$ where ${\cal L}_{M}$ thus depends on
$x$ only
through the fields.

The usual conception of Poincar$\acute{\mbox{e}}$ symmetry partly as a
spacetime
partly as an inner symmetry is expressed in the transformation formulae
\beqq \label{6} x^{\al}\lar
x'^{\al}&=&x^{\al}+\ep^{\al}+\om^{\al}\,_{\be}x^{\be}, \\
\va_{j}(x)\lar
\va_{j}'(x')&=&\va_{j}(x)-\frac{i}{4}\om^{\al\be}\Si_{\al\be}\,\,\va_{j}(x)
.\nonumber \eeqq
$\de x^{\al}=\ep^{\al}+\om^{\al}\,_{\be}x^{\be}$ is the change of $x$ under the
combination of a global infinitesimal spacetime translation
and a global infinitesimal Lorentz rotation,
$\de\va_{j}=-\frac{i}{4}\om^{\al\be}\Si_{\al\be}\,\,\va_{j}(x)$ the
corresponding change of $\va_{j}$
in field space. $\Si_{\ga\de}$ are the representations of the generators of the
Lie
algebra {\bf so(1,3)\/} in inner field space normalized to fulfil the
commutation
relations
\beq
[\Si_{\ga\de},\Si_{\ep\zeta}]=2i\{\eta_{\de\ep}\Si_{\ga\zeta}-\eta_{\de\zeta}
\Si_{\ga\ep}+\eta_{\ga\ep}\Si_{\zeta\de}-\eta_{\ga\zeta}\Si_{\ep\de}\}. \eeq
One easily convinces oneself now that eqn.(\ref{2}) holds for
$\de f^{\ga}=0$ and obtains with (\ref{3}) the conserved current
\beq \label{8} J^{\ga}=\Theta^{\ga}\,_{\al}\cdot\ep^{\al}+\frac{1}{2}{\cal
M}^{\ga}
\,_{\al\be}\cdot\om^{\al\be} \eeq
where
${\cal M}^{\ga}\,_{\al\be}$ is the canonical angular momentum tensor
\beq {\cal M}^{\ga}\,_{\al\be}=\Theta^{\ga}\,_{\al}x_{\be}-\Theta^{\ga}\,_{\be}
x_{\al}+\frac{i}{2}\frac{\pa{\cal L}_{M}}{\pa(\pa_{\ga}\va_{j})}
\Si_{\al\be}\,\va_{j}. \eeq

Next, we introduce global infinitesimal {\bf P\/} gauge transformations
\beqq \label{10} x^{\al}\lar x'^{\al}&=&x^{\al}, \nonumber \\ \va_{j}(x)\lar
\va_{j}'(x)
&=&\va_{j}(x)-\{\ep^{\al}+\om^{\al}\,_{\be}x^{\be}\}\cdot\pa_{\al}\va_{j}(x) \\
&-&\frac{i}{4}\om^{\al\be}\Si_{\al\be}\,\,\va_{j}(x).\nonumber \eeqq
{\bf P\/} acts now as a group of generalized 'phase rotations' in field space
only
and leaves the spacetime coordinates $x$ unchanged. As it is again a symmetry
transformation of Poincar$\acute{\mbox{e}}$ invariant actions we are led to the
complementary conception of Poincar$\acute{\mbox{e}}$ symmetry as a purely
inner
symmetry. We now have $\de x^{\al}=0$,
$\de\va_{j}=-\{\ep^{\al}+\om^{\al}\,_{\be}x^{\be}\}
\cdot\pa_{\al}\va_{j}-\frac{i}{4}\om^{\al\be}\Si_{\al\be}\,\,\va_{j}$ and
\beqq & &\quad d^{4}x\,{\cal L}_{M}(\va_{j}'(x),\pa_{\al}\va_{j}'(x))= \\
& &d^{4}x\,\{{\cal L}_{M}(\va_{j}(x),\pa_{\al}\va_{j}(x))-
\{\ep^{\ga}+\om^{\ga}\,_{\be}x^{\be}\}\cdot\pa_{\ga}
{\cal L}_{M}(\va_{j}(x),\pa_{\al}\va_{j}(x))\} \nonumber \eeqq
so that eqn.(\ref{2}) holds with $\de
f^{\ga}=-\delta^{\ga}\,_{\al}\cdot\{\ep^{\al}+
\om^{\al}\,_{\be}x^{\be}\}{\cal L}_{M}$. The conserved current is found to be
the same
$J^{\ga}$ as in eqn.(\ref{8}). This shows that the two complementary
conceptions are
equivalent w.r.t. their physical consequences. In both cases the conserved
charges
are found to be the energy-momentum
\beq P_{\al}=\int d^{3}x\,\Theta^{0}\,_{\al} \eeq
and the angular momentum
\beq M_{\al\be}=\int d^{3}x\,{\cal M}^{0}\,_{\al\be}. \eeq

We have obtained a description of the global action of {\bf P\/} which
resembles
very much the manner the action of well-known inner symmetries is usually
described in
field theory (see e.g. \cite{lor}). Let us now go one step further and gauge
the
Poincar$\acute{\mbox{e}}$ group {\bf P\/} extending the discussion of gauging
{\bf T\/} in \cite{wie}.

\section{Local P gauge invariance. The covariant derivative
${\tilde\nabla}_{\al}$ and
its decomposition w.r.t. $p_{\ga}$ and $m_{\ga\de}$}

\paragraph{}
In this section we introduce local {\bf P\/} gauge transformations and the
corresponding covariant derivative ${\tilde\nabla}_{\al}=\pa_{\al}+B_{\al}$
respecting the
local {\bf P\/} gauge symmetry. We give the decomposition of the compensating
field
$B_{\al}$ w.r.t. the {\bf p\/} generators $p_{\ga}$ and $m_{\ga\de}$ and
determine its
behaviour under local {\bf P\/} gauge transformations.

In the previous section we recast {\bf P\/} symmetry in the form of an inner
symmetry.
Only in this conception it is possible to rewrite the formulae (\ref{10}) for
global
infinitesimal {\bf P\/} gauge transformations as
\beqq \label{14} x^{\al}\lar x'^{\al}&=&x^{\al} \\ \va_{j}(x)\lar
\va_{j}'(x)&=&
\Bigl(({\bf 1}+\Theta)\va_{j}\Bigr)(x). \nonumber \eeqq
Hence, we can introduce in complete analogy to notions used in non-abelian
gauge field
theory the unitary infinitesimal representations of {\bf P\/} transformations
in field
space
$({\bf 1}+\Theta)$ forming a Lie group, where the infinitesimal hermitean gauge
operators
\beqq \Theta&=&-\{\ep^{\ga}+\om^{\ga\de}x_{\de}\}\cdot\pa_{\ga}
-\frac{i}{4}\om^{\ga\de}\Si_{\ga\de} \\ &=& i\,\ep^{\ga}\cdot p_{\ga}
-\frac{i}{2}\om^{\ga\de}\cdot m_{\ga\de} \nonumber \eeqq
are decomposed in two equivalent ways for later use. The corresponding
generators of
the Lie algebra {\bf p\/} in field space
\beq p_{\ga}=i\,\pa_{\ga}, \quad\quad
m_{\ga\de}=i(x_{\ga}\pa_{\de}-x_{\de}\pa_{\ga})
+\frac{1}{2}\Si_{\ga\de} \eeq
are normalized to fulfil the usual commutation relations of the {\bf p\/}
generators
\beqq \label{17} \lbr p_{\ga},p_{\de}\rbr&=&0 \nonumber \\
\lbr p_{\ga},m_{\ep\ze}\rbr &=&i\{\eta_{\ga\ep}p_{\ze}-\eta_{\ga\ze}p_{\ep}\}
\\
\lbr m_{\ga\de},m_{\ep\ze}\rbr
&=&i\{\eta_{\de\ep}m_{\ga\ze}-\eta_{\de\ze}m_{\ga\ep}
+\eta_{\ga\ep}m_{\ze\de}-\eta_{\ga\ze}m_{\ep\de}\}. \nonumber \eeqq
Above hermiticity and unitarity are understood w.r.t. the usual scalar product
in
field space.

Let us extend now {\bf P\/} to a Lie group of local infinitesimal
gauge transformations allowing $\ep(x)$ and $\om(x)$ to vary with $x$. We thus
consider from now on
\beqq \label{18}
\Theta(x)&=&-\{\ep^{\ga}(x)+\om^{\ga\de}(x)x_{\de}\}\cdot\pa_{\ga}
-\frac{i}{4}\om^{\ga\de}(x)\Si_{\ga\de} \\ &=& i\,\ep^{\ga}(x)\cdot p_{\ga}
-\frac{i}{2}\om^{\ga\de}(x)\cdot m_{\ga\de}. \nonumber \eeqq
Note that the algebra of the $\Theta(x)$ does close again. There is a new
element of
non-commutativity in the algebra of the $\Theta$ besides the one expressed in
(\ref{17}) as, contrary to the usual case, the local parameters $\ep(x)$ and
$\om(x)$
don't commute with the generators of the algebra given in (\ref{17}). The
emerging
ordering problem is overcome by the convention that $\Theta(x)$ in its above
form only
acts to the right. This convention is motivated by demanding equivalence of the
algebra
of the $\Theta(x)$ to the diffeomorphism times {\bf so(1,3)\/} algebra. The
formulae
(\ref{14}) still define the representation of {\bf P\/} in the space of fields.

In order to recast a given matter theory in a locally {\bf P\/} gauge invariant
form
we must introduce a covariant derivative ${\tilde\nabla}_{\al}$. To be more
precise we
demand that the Lagrangian with covariant derivatives ${\tilde\nabla}_{\al}$
replacing the
usual ones behave under local infinitesimal {\bf P\/} transformations the same
way it
behaves with the usual derivatives $\pa_{\al}$ under global infinitesimal {\bf
P\/}
transformations. Hence, we must ensure
\beqq \label{19} & &\quad{\cal
L}_{M}(\va_{j}'(x),{\tilde\nabla}'_{\al}\va_{j}'(x))= \\
& &{\cal L}_{M}(\va_{j}(x),{\tilde\nabla}_{\al}\va_{j}(x))-
\{\ep^{\ga}(x)+\om^{\ga\de}(x)x_{\de}\}\cdot\pa_{\ga}
{\cal L}_{M}(\va_{j}(x),{\tilde\nabla}_{\al}\va_{j}(x)) \nonumber \eeqq
where ${\tilde\nabla}'_{\al}$ denotes the gauge transformed covariant
derivative. Note
that (\ref{19}) alone does not lead to the local {\bf P\/} invariance of the
original
action $S_{M}=\int{\cal L}_{M}$ for the second term in eqn.(\ref{19}) is no
longer a
pure divergence as it was in the case of global infinitesimal transformations.

As usual it is sufficient to construct a covariant derivative which fulfils
\beq \label{20} {\tilde\nabla}'_{\al}\,\left({\bf
1}+\Theta(x)\right)=\left({\bf 1}
+\Theta(x)\right)\,{\tilde\nabla}_{\al}. \eeq
Because ${\tilde\nabla}_{\al}$ transforms as a Lorentz vector we have to
supplement the generators $\Si_{\ga\de}$ of {\bf so(1,3)\/} in matter field
space
occuring in the decomposition of $\Theta(x)$ with the corresponding generators
$\Si_{\ga\de}$ acting on vectors to obtain the appropriate product
representation as
we will always do where necessary.
For the infinitesimal transformations considered, eqn.(\ref{20}) ensures indeed
the
proper transformation behaviour (\ref{19}). As $\Theta(x)$ in eqn.(\ref{18})
may
be decomposed w.r.t. the Lie algebra generators as usual we are led to try the
ansatz
\beq \label{21} \pa_{\al}\lar {\tilde\nabla}_{\al}=\pa_{\al}+B_{\al} \eeq
together with the decomposition of $B_{\al}$ w.r.t. $\pga$ and $\mga$
\beq \label{22} B_{\al}\equiv -i\,\bga\cdot\pga+\frac{i}{2}\bbga\cdot\mga \eeq
introducing the 40 compensating fields $\bga$ and $\bbga$. We emphasize that
this
decomposition, relevant for any perturbative calculation, yields as fundamental
compensating fields the 16 $\bga$ for the local translations and the 24 $\bbga$
for
the local Lorentz rotations. It is only possible in the context of gauging the
Poincar$\acute{\mbox{e}}$ group as an inner symmetry group. The spin generators
$\Si_{\ga\de}$ occuring in the decomposition of $B_{\al}$ always have to be
adjusted to
the Lorentz group representation upon which they act, hence manifestly
ensuring the covariant transformation behaviour of ${\tilde\nabla}_{\al}$
throughout.
Note that $B_{\al}$ acts here not only as a matrix but also as a differential
operator
in field space.

We are ready now to discuss the behaviour of $B_{\al}$ under local gauge
transformations. Inserting the ansatz (\ref{21}) for ${\tilde\nabla}_{\al}$ in
eqn.(\ref{20}) we obtain the transformation law
\beqq \label{23} \de B_{\al}&=&B'_{\al}-B_{\al}=[\Theta,\pa_{\al}+B_{\al}]\\
&=&-\pa_{\al}\Theta+\om_{\al}\,^{\be}\pa_{\be}+[\Theta,B_{\al}] \nonumber \eeqq
where we remark that the second term in (\ref{23}) just cancels the last term
in
\bed \pa_{\al}\Theta=i\,\pa_{\al}\ep^{\ga}\cdot p_{\ga}-\frac{i}{2}\pa_{\al}
\om^{\ga\de}\cdot m_{\ga\de}-\frac{i}{2}\om^{\ga\de}\cdot\pa_{\al}m_{\ga\de}.
\eed
Hence, there are only the derivatives of $\ep(x)$ and $\om(x)$ occuring above
as expected.
Eqn.(\ref{23}) defines the representation of the local
Poincar$\acute{\mbox{e}}$ group
{\bf P\/} in the gauge field space. Note the structural similarity of the
results
obtained up to now to similar ones in the discussion of non-abelian gauge
symmetry
\cite{lor}.
Next we decompose $\de B_{\al}$ w.r.t. the generators $\pga$ and $\mga$ of {\bf
p\/}
\beq \de B_{\al}\equiv -i\,\de\bga\cdot\pga+\frac{i}{2}\de\bbga\cdot\mga. \eeq
The quite lengthy evaluation of all the commutators shows that $\de B_{\al}$
has indeed
the required decomposition and we obtain
\beqq \label{25}
\de\bga&=&\pa_{\al}\ep^{\ga}+B_{\al}\,^{\ze}\cdot\pa_{\ze}\ep^{\ga}
-\bbep\,x_{\ep}\cdot\pa_{\ze}\ep^{\ga}-\ep^{\ze}\cdot\pa_{\ze}\bga \\
&+&\om^{\ep\ze}\,x_{\ep}\cdot\pa_{\ze}\bga-\ep_{\de}\bbga+\om_{\al}\,^{\be}
B_{\be}\,^{\ga}+\om^{\ga}\,_{\de}B_{\al}\,^{\de} \nonumber \eeqq
and
\beqq \label{26}
\de\bbga&=&\pa_{\al}\om^{\ga\de}+B_{\al}\,^{\ze}\cdot\pa_{\ze}\om^{\ga\de}
-\bbep\,x_{\ep}\cdot\pa_{\ze}\om^{\ga\de}-\ep^{\ze}\cdot\pa_{\ze}\bbga \\
&+&\om^{\ep\ze}\,x_{\ep}\cdot\pa_{\ze}\bbga+\om_{\al}\,^{\be}B_{\be}\,^{\ga\de}
+\om^{\ga}\,_{\ze}B_{\al}\,^{\ze\de}+\om^{\de}\,_{\ze}B_{\al}\,^{\ga\ze}
. \nonumber \eeqq
Compared with the corresponding transformation formula in the translation gauge
invariant
theory \cite{wie} eqns.(\ref{25}) and (\ref{26}) become quite involved and
again
strongly differ from the analogous ones in non-abelian gauge field theory.

\section{The covariant derivative ${\tilde\nabla}_{\al}$ and
its decomposition w.r.t. $\pa_{\ga}$ and $\Si_{\ga\de}$}

\paragraph{}
In this section we decompose the covariant derivative ${\tilde\nabla}_{\al}$
w.r.t.
$\pa_{\ga}$ and $\Si_{\ga\de}$ and recast it in terms of the effective gauge
fields
$\ega$ and $\bbga$. We introduce the field strength operator and determine its
behaviour under local {\bf P\/} gauge transformations.

The decomposition of the covariant derivative ${\tilde\nabla}_{\al}$ given in
the
previous section emphasizes the relation of the fundamental compensating fields
to the
Poincar$\acute{\mbox{e}}$ algebra {\bf p\/} and is a crucial tool for all
perturbative
calculations in the present approach. To obtain the covariant
objects of the theory in a compact form, however, it is suitable to recast
${\tilde\nabla}_{\al}$ from eqns.(\ref{21}), (\ref{22})
\beq {\tilde\nabla}_{\al}=\ega\pa_{\ga}+\frac{i}{4}\bbga\Si_{\ga\de} \eeq
introducing the effective matrix fields $\ega$
\beq \ega\equiv\dga+\bga+\bbga x_{\de}. \eeq
Note that this decomposition of ${\tilde\nabla}_{\al}$ corresponds just to the
first
way of expressing the local gauge operator $\Theta$ in eqn.(\ref{18}).
Abbreviating
\beq d_{\al}\equiv\ega\pa_{\ga},\quad\quad
B_{\al}\equiv\frac{i}{4}\bbga\Si_{\ga\de},\eeq
where $\Si_{\ga\de}$ must be properly adjusted to the Lorentz group
representation
it acts upon, we write ${\tilde\nabla}_{\al}=d_{\al}+B_{\al}$ from now on.
$d_{\al}$
is just the translation covariant derivative introduced in \cite{wie}.

As in our conception coordinate and {\bf P\/} gauge transformations are
strictly
separated we emphasize that the introduction of $\bga$, $\bbga$ and $\ega$ has
neither
implications on the structure of the underlying spacetime which we assumed to
be
({\bf R\/}$^{4}$,$\eta$) endowed with the Minkowski metric $\eta$. Nor has it
implications on the maximal symmetry group of ({\bf R\/}$^{4}$,$\eta$), which
is the
Poincar$\acute{\mbox{e}}$ group if we still restrict ourselves to the use of
Cartesian
coordinates only. This fact will allow one to obtain the energy-momentum and
angular
momentum of the gauge fields by an application of the Noether theorem given in
section
1.

Let us now recast the quite involved transformation behaviour of $\bga$ and
$\bbga$
under local {\bf P\/} gauge transformations in terms of $\ega$ and $\bbga$.
With the
use of eqns.(\ref{25}) and (\ref{26}) the variation of $\ega$ becomes quite
simple
\beqq \label{30} \de\ega
&=&e_{\al}\,^{\ze}\cdot\pa_{\ze}\{\ep^{\ga}+\om^{\ga\de}x_{\de}\} \\
&-&\{\ep^{\ze}+\om^{\ze\eta}x_{\eta}\}\cdot\pa_{\ze}\ega
+\om_{\al}\,^{\ze}e_{\ze}\,^{\ga} \nonumber \eeqq
and is expressed in terms of $\ega$ only. For the variation of
$\bbga$ we obtain the result
\beqq \label{31} \de\bbga&=&e_{\al}\,^{\ze}\cdot\pa_{\ze}\om^{\ga\de}
-\{\ep^{\ze}+\om^{\ze\eta}x_{\eta}\}\cdot\pa_{\ze}\bbga \\
&+&\om_{\al}\,^{\ze}B_{\ze}\,^{\ga\de}+\om^{\ga}\,_{\ze}B_{\al}\,^{\ze\de}
+\om^{\de}\,_{\ze}B_{\al}\,^{\ga\ze}. \nonumber \eeqq
As the determinant $\dem$ will enter the locally {\bf P\/} invariant actions
we give its transformation behaviour already here
\beqq \label{32}
\de\dem&=&-\dem\cdot\pa_{\ze}\{\ep^{\ze}+\om^{\ze\eta}x_{\eta}\} \\
&-&\{\ep^{\ze}+\om^{\ze\eta}x_{\eta}\}\cdot\pa_{\ze}\dem. \nonumber \eeqq

Before turning to the field strength operator we introduce the non-covariant
decomposition
\beq [d_{\al},d_{\be}]\equiv\hga d_{\ga} \eeq
as in \cite{wie}. $\hga$ is expressed in terms of $\ega$ as
\beq \hga=e^{-1\,\ga}\,_{\ep}(e_{\al}\,^{\ze}\cdot\pa_{\ze} e_{\be}\,^{\ep}-
e_{\be}\,^{\ze}\cdot\pa_{\ze} e_{\al}\,^{\ep}) \eeq
where $e^{-1\,\ga}\,_{\ep}$ is the matrix inverse to $e_{\al}\,^{\ep}$, i.e.
$e_{\al}\,^{\ep}\cdot e^{-1\,\ga}\,_{\ep}=\de_{\al}\,^{\ga}$.

This allows us now to obtain the field strength operator and its decomposition.
Taking into account the vector character of ${\tilde\nabla}_{\al}$ we obtain
after a
little algebra
\beqq S_{\al\be}&\equiv&[{\tilde\nabla}_{\al},{\tilde\nabla}_{\be}] \nonumber
\\
&=&\hga d_{\ga}-(B_{\al\be}\,^{\ga}-B_{\be\al}\,^{\ga})d_{\ga} \\
&+&d_{\al}B_{\be}-d_{\be}B_{\al}+[B_{\al},B_{\be}]. \nonumber \eeqq
Introducing the tensor coefficients of $d_{\ga}$
\beq \label{36} T_{\al\be}\,^{\ga}\equiv B_{\al\be}\,^{\ga}-B_{\be\al}\,^{\ga}
-H_{\al\be}\,^{\ga} \eeq
we may rewrite $S_{\al\be}$ as
\beq \label{37}
[{\tilde\nabla}_{\al},{\tilde\nabla}_{\be}]=-T_{\al\be}\,^{\ga}\,
{\tilde\nabla}_{\ga}+\frac{i}{4}{\tilde R}^{\ga\de}\,_{\al\be}\Si_{\ga\de},
\eeq
where ${\tilde R}^{\ga\de}\,_{\al\be}$ is found to be
\beqq \label{38} {\tilde R}^{\ga\de}\,_{\al\be}&\equiv&
d_{\al}B_{\be}\,^{\ga\de}-d_{\be}B_{\al}\,^{\ga\de}+B_{\al}\,^{\de\ep}
B_{\be\ep}\,^{\ga} \\ &-&B_{\be}\,^{\de\ep} B_{\al\ep}\,^{\ga}
-H_{\al\be}\,^{\ep} B_{\ep}\,^{\ga\de}. \nonumber \eeqq
For later use we finally introduce the shorthand notation
\beq \label{39} {\tilde R}\,_{\al\be}\equiv\frac{i}{4}{\tilde
R}^{\ga\de}\,_{\al\be}
\Si_{\ga\de}. \eeq

As $S_{\al\be}$ has a decomposition w.r.t. ${\tilde\nabla}_{\de}$ and
$\Si_{\ga\de}$
it acts in general not only as a matrix but also as a first order differential
operator in field space. Only if $\bbga$ is related to $H_{\al\be}\,^{\ga}$ the
coefficient $T_{\al\be}\,^{\ga}$
of the operator part in eqn.(\ref{37}) does vanish. Denoting this particular
choice
of $\bbga$ being of much importance later with $\ccga$ the required relation
becomes
\beq C_{\al\be}\,^{\ga}-C_{\be\al}\,^{\ga}=H_{\al\be}\,^{\ga}. \eeq
We may now solve for $\ccga$ with the result
\beq \ccga=\frac{1}{2}\left(H_{\al}\,^{\ga\de}-H_{\al}\,^{\de\ga}
-H^{\ga\de}\,_{\al}\right). \eeq
For the special choice $\bbga=\ccga$ we omit the tilde, hence writing
\beq \label{42} \nabla_{\al}\equiv d_{\al}+C_{\al}. \eeq
Obviously we obtain for $S_{\al\be}$ a matrix only
\beq \label{43} [\nabla_{\al},\nabla_{\be}]=\frac{i}{4}R^{\ga\de}\,_{\al\be}
\Si_{\ga\de}\equiv R\,_{\al\be} \eeq
where
\beqq R^{\ga\de}\,_{\al\be}&=&d_{\al}C_{\be}\,^{\ga\de}
-d_{\be}C_{\al}\,^{\ga\de}+C_{\al}\,^{\de\ep}C_{\be\ep}\,^{\ga} \\
&-&C_{\be}\,^{\de\ep}C_{\al\ep}\,^{\ga}
-(C_{\al\be}\,^{\ep}-C_{\be\al}\,^{\ep}) C_{\ep}\,^{\ga\de} \nonumber \eeqq
is now expressed in terms of $\ccga$.

By construction $S_{\al\be}$ transforms homogeneously under infinitesimal local
{\bf P\/} gauge transformations
\beq S'_{\al\be}\left({\bf 1}+\Theta(x)\right)=\left({\bf 1}+\Theta(x)\right)
\,S_{\al\be} \eeq
and thus
\beq \label{45} \de S_{\al\be}=S'_{\al\be}-S_{\al\be}=[\Theta,S_{\al\be}] \eeq
Decomposition of eqn.(\ref{45}) w.r.t. ${\tilde\nabla}_{\ga}$ and
$\Si_{\ga\de}$
together with the use of the known transformation law eqn.(\ref{30}) for $\ega$
leads
to
\beqq \de
T_{\al\be}\,^{\ga}&=&-\{\ep^{\ze}+\om^{\ze\eta}x_{\eta}\}\cdot\pa_{\ze}
T_{\al\be}\,^{\ga} \\
&+&\om_{\al}\,^{\ze}T_{\ze\be}\,^{\ga}+\om_{\be}\,^{\ze}T_{\al\ze}\,^{\ga}
+\om^{\ga}\,_{\ze}T_{\al\be}\,^{\ze} \nonumber \eeqq
and to
\beqq \de {\tilde
R}^{\ga\de}\,_{\al\be}&=&-\{\ep^{\ze}+\om^{\ze\eta}x_{\eta}\}\cdot
\pa_{\ze} {\tilde R}^{\ga\de}\,_{\al\be}+\om_{\al}\,^{\ze}{\tilde
R}^{\ga\de}\,_{\ze\be} \\
&+&\om_{\be}\,^{\ze}{\tilde R}^{\ga\de}\,_{\al\ze}+\om^{\ga}\,_{\ze}
{\tilde R}^{\ze\de}\,_{\al\be}+\om^{\de}\,_{\ze}{\tilde R}^{\ga\ze}\,_{\al\be}.
\nonumber \eeqq
$T_{\al\be}\,^{\ga}$ and ${\tilde R}^{\ga\de}\,_{\al\be}$ transform
homogeneously
under infinitesimal local {\bf P\/} gauge transformations. We emphasize
that the choice $T_{\al\be}\,^{\ga}=0$ is indeed a gauge covariant statement as
we
implicitly assumed above introducing $\ccga$. As long as one works with
regularizations
respecting the gauge symmetry, as we will do later on, it is always possible to
work
consistently under the constraint $T=0$.

For later use we finally introduce the difference of the two gauge fields
\beq \label{49} K_{\al}\,^{\ga\de}\equiv \bbga-\ccga \eeq
which is related to $T_{\al\be}\,^{\ga}$ as
\beq K_{\al\be\ga}-K_{\be\al\ga}=T_{\al\be\ga} \eeq
with the obvious inversion
\beq K_{\al\ga\de}=\frac{1}{2}\left(T_{\al\ga\de}-T_{\al\de\ga}
-T_{\ga\de\al}\right). \eeq

\section{{\bf P\/} gauge invariant matter actions. Scalar, spinor and vector
fields
as examples}

\paragraph{}
In this section we discuss the extension of globally {\bf P\/} invariant matter
actions on ({\bf R\/}$^{4}$,$\eta$) to locally {\bf P\/} gauge invariant ones.
We then
apply the general framework to a scalar, spinor and vector field action in turn
and determine their respective locally {\bf P\/} gauge invariant forms.

Let us consider a globally {\bf P\/} invariant theory for $n$ fields $\va_{j}$
specified
by the Lagrangian density ${\cal L}_{M}(\va_{j},\pa_{\al}\va_{j})$ assumed to
be
real ${\cal L}^{\ast}_{M}={\cal L}_{M}$. In section 3 we have constructed a
covariant
derivative ${\tilde\nabla}_{\al}$ which respects the behaviour of ${\cal
L}_{M}$
under global Poincar$\acute{\mbox{e}}$ gauge transformations extended to local
ones
as expressed in eqn.(\ref{19}). But as we already mentioned (\ref{19}) is not
yet
sufficient for the original action $S_{M}=\int {\cal L}_{M}$ to be locally {\bf
P\/}
gauge invariant.

We have to complete the Lagrangian density with another term ensuring that the
change
of both parts together under a local {\bf P\/} transformation will yield a pure
divergence only. Using the transformation law (\ref{32}) for $\dem$ we get for
the behaviour of the combination
\beq \label{52} \dem\cdot{\cal L}_{M}(\va_{j},{\tilde\nabla}_{\al}\va_{j}) \eeq
under local {\bf P\/} gauge transformations
\beqq & &\det e'^{-1}\cdot{\cal L}_{M}(\va_{j}',{\tilde\nabla}'_{\al}\va_{j}')=
\dem\cdot{\cal L}_{M}(\va_{j},{\tilde\nabla}_{\al}\va_{j}) \\ & &\quad\quad
-\pa_{\ga}\left(\{\ep^{\ga}(x)+\om^{\ga\de}(x)x_{\de}\}\dem\cdot
{\cal L}_{M}(\va_{j},{\tilde\nabla}_{\al}\va_{j})\right), \nonumber \eeqq
i.e. the change of the combination (\ref{52}) is indeed a pure divergence.

Therefore the minimally extended locally {\bf P\/} gauge invariant matter
action
becomes
\beq \label{56} S_{M}=\intf \dem(x)\cdot{\cal L}_{M}(\va_{j}(x),
{\tilde\nabla}_{\al}\va_{j}(x)). \eeq
Of course, $S_{M}$ remains invariant if we change from one to another inertial
system
by global coordinate translations or Lorentz rotations.

It is the conception of {\bf P\/} symmetry as an inner symmetry together with
the
gauge principle which has led us to this minimal coupling prescription. In this
conception the gauge fields and their transformation behaviour do not interfere
with the spacetime structure ({\bf R\/}$^{4}$,$\eta$) fixed by an a priori
convention
and the underlying geometry remains separated from the physics described by the
{\bf P\/}
gauge fields in the same manner it remains separated from the physics described
by
any usual matrix gauge field.

For later use we turn now to apply the general framework developed so far to a
real
massive scalar field, a massive Dirac spinor and a massive vector field. The
globally
{\bf P\/} invariant action for the scalar field $\va$ is given by
\beq \label{55} S_{M}=\intf\left\{ \frac{1}{2}\,\pa_{\al}\va\cdot\pa^{\al}\va
-\frac{1}{2}m^{2}\va^{2}\right\}. \eeq
The {\bf so(1,3)\/} generators are trivial and the {\bf P\/} covariant
derivative
is independent of $\bbga$ as is the minimal extension of (\ref{55}) to a
locally
{\bf P\/} gauge invariant action
\beq S_{M}=\intf\dem\left\{ \frac{1}{2} d_{\al}\va\cdot d^{\al}\va
-\frac{1}{2}m^{2}\va^{2}\right\}. \eeq
We obtain the same result as in the case of pure {\bf T\/} gauge invariance
\cite{wie}.
Note that in the presence of $\ega$ the scalar product in real scalar field
space
becomes now $\left(\chi,\va\right)_{e}=\intf\dem\,\chi\cdot\va$.

The globally {\bf P\/} invariant action for a Dirac spinor with real Lagrangian
density
is given by
\beq S_{M}=\intf\left\{ \frac{i}{2}\psiba\ga^{\al}(\pa_{\al}\psi)-\frac{i}{2}
\overline{(\pa_{\al}\psi)}\ga^{\al}\psi-m\psiba\psi\right\}. \eeq
The Dirac matrices fulfil the usual Clifford algebra
$\{\ga_{\al},\ga_{\be}\}=2\eta_{\al\be}$ and the {\bf so(1,3)\/} generators
become
$\Si_{\al\be}=\frac{i}{2}[\ga_{\al},\ga_{\be}]$.
The minimal extension prescription yields the locally {\bf P\/} gauge invariant
action
\beq S_{M}=\intf\dem\left\{
\frac{i}{2}\psiba\ga^{\al}({\tilde\nabla}_{\al}\psi)-\frac{i}{2}
\overline{({\tilde\nabla}_{\al}\psi)}\ga^{\al}\psi-m\psiba\psi\right\}. \eeq
Due to spin $\bbga$ enters now the action. We will further investigate this
below in
the context of quantum field theoretical considerations. Partially integrating
${\tilde\nabla}_{\al}$ in the second term above leads to the usual form of the
Dirac action
\beq \label{59} S_{M}=\intf\dem\psiba\left\{ i\ga^{\al}({\tilde\nabla}_{\al}
-\frac{1}{2}K_{\ga\al}\,^{\ga})-m\right\}\psi. \eeq
Note the occurrence of the tensor $K$ ensuring the hermiticity of the {\bf P\/}
covariant Dirac operator w.r.t.
$\left(\chi,\psi\right)_{e}=\intf\dem\,\overline{\chi}\cdot\psi$.

We turn to the last example. The globally {\bf P\/} invariant action for a
massive
vector field is given by
\beq \label{60} S_{M}=\intf\left\{-\frac{1}{4}F_{\al\be}F^{\al\be}
+\frac{1}{2}m^{2}A_{\al}A^{\al}\right\}, \eeq
where the field strength reads $F_{\al\be}=\pa_{\al}A_{\be}-\pa_{\be}A_{\al}$.
The
{\bf so(1,3)\/} generators in the vector representation are
$(\Si_{\al\be})^{\ga\de}=2i(\eta_{\al}\,^{\ga}\eta_{\be}\,^{\de}
-\eta_{\al}\,^{\de}\eta_{\be}\,^{\ga})$
and the minimal extension of the action (\ref{60}) to a locally {\bf P\/} gauge
invariant action yields
\beq \label{61} S_{M}=\intf\dem\left\{-\frac{1}{4}F_{\al\be}F^{\al\be}
+\frac{1}{2}m^{2}A_{\al}A^{\al}\right\} \eeq
together with the covariant field strength
$F_{\al\be}={\tilde\nabla}_{\al}A_{\be}-{\tilde\nabla}_{\be}A_{\al}$.
The covariant derivative in the vector representation is simply
${\tilde\nabla}_{\al}A_{\be}=d_{\al}A_{\be}-B_{\al\be}\,^{\ga}A_{\ga}$.
We remark that under the {\bf P\/} gauge covariant $U(1)$ gauge transformation
\bed A_{\al}\lar A_{\al}+\nabla_{\al}\La \eed
the field strength is in general not invariant
\bed F_{\al\be}\lar F_{\al\be}-T_{\al\be}\,^{\ga}\nabla_{\ga}\La. \eed
Only for $T=0$ the Maxwell action with $m=0$ is $U(1)$ gauge invariant. The
scalar
product in vector field space
$\left(A,B\right)_{e}=-\intf\dem\,A_{\al}\eta^{\al\be}B_{\be}$
is chosen such that the physical polarizations have positive norm.

All the examples above and non-minimal extensions are discussed in the
geometrical
framework e.g. in \cite{buc}.

\section{Invariance under coordinate transformations and frame rotations as
complementary conception}

\paragraph{}
In this section the concept of local {\bf P\/} gauge invariance is shown to be
physically equivalent to the usual concepts of coordinate and local Lorentz
invariance.
This equivalence allows us to re-interpret the formalism in common geometrical
terms.

Up to now we have relied on the conception of {\bf P\/} symmetry as an inner
symmetry
expressed in the transformation behaviour eqn.(\ref{14}) for matter fields and
eqns.
(\ref{30}), (\ref{31}) for the gauge fields. It allowed us to extend the
framework of
gauge theories of matrix groups to the operator gauge group {\bf P\/}. As the
Poincar$\acute{\mbox{e}}$ group of global spacetime transformations relating
different observers and the local gauge group {\bf P\/} were
strictly separated the a priori geometry of spacetime ({\bf R\/}$^{4}$,$\eta$)
chosen to be Minkowskian was not affected by the introduction of the {\bf P\/}
gauge
fields $\ega$ and $\bbga$.

We turn now to the complementary conception of Poincar$\acute{\mbox{e}}$
symmetry
partly as a spacetime partly as an inner symmetry (\cite{kib}-\cite{hen}) and
introduce
besides the
orthonormal indices $\al,\be,\ga,...$ used up to now the coordinate indices
$\mu,\nu,\rho,...\,$. The infinitesimal transformation formulae involve now
both
coordinates and fields as in eqn.(\ref{6})
\beqq x^{\mu}\lar x'^{\mu}&=&x^{\mu}+\ep^{\mu}(x), \\
\va_{j}(x)\lar \va_{j}'(x')&=&\va_{j}(x)
-\frac{i}{4}\om^{\al\be}(x)\Si_{\al\be}\,\,\va_{j}(x).\nonumber \eeqq
Note that $\ep^{\mu}(x)$ parametrizes a general infinitesimal coordinate
transformation
and may contain effects of local translations as well as local Lorentz
rotations of the
coordinates which may actually no longer be distinguished
(\cite{kib}-\cite{hen}).
$\om^{\al\be}(x)
=-\om^{\be\al}(x)$ parametrizes now a local orthonormal frame rotation and is
in no
respect related to $\ep^{\mu}(x)$. This becomes manifest if we rewrite
eqn.(\ref{30})
in the equivalent form
\beq \emu(x)\lar e'_{\al}\,^{\mu}(x')=(\de_{\al}\,^{\be}+\om_{\al}\,^{\be})
e_{\be}\,^{\nu}(x)(\de_{\nu}\,^{\mu}+\pa_{\nu}\ep^{\mu}). \eeq
We find that coordinate indices $\mu,\nu,\rho,...$ are transformed with the
Jacobi
matrix $\frac{\pa x'^{\mu}}{\pa x^{\nu}}=\de_{\nu}\,^{\mu}+\pa_{\nu}\ep^{\mu}$
resulting from the infinitesimal coordinate change and orthonormal indices with
the
infinitesimal frame rotation $\de_{\al}\,^{\be}+\om_{\al}\,^{\be}$.

Hence, the gauge group {\bf P\/} and the requirement of local {\bf P\/} gauge
invariance
are replaced by the groups of general (infinitesimal) coordinate
transformations and
local $SO(1,3)$ frame rotations and the requirement of invariance under these
groups
\cite{sexu}, \cite{str}.
Indeed, in many other gauge approaches to gravitation some combination of these
two
groups is used as the gauge group \cite{kib}-\cite{hen}.

As a consequence $e_{\al}\,^{\mu}$ has to be re-interpreted as vierbein and
defines a
metric tensor
\beq \label{64} \gmu=e_{\al}\,^{\mu}\,e^{\al\nu}. \eeq
The geometry of spacetime is now necessarily linked with the above discussed
complementary symmetry requirements and Riemannian geometry becomes
the natural framework to deal with this point of view. The
geometric notations used here correspond to those in \cite{naka}. $\pa_{\mu}$
and
$\hat{e}_{\al}\equiv d_{\al}$ fit in as coordinate and orthonormal
non-coordinate basis
vectors in the tangential spaces belonging to the Riemannian manifold
({\bf R\/}$^{4}$,$g$). The manifold is  endowed with the indefinite metric
$\gmu$
defined in eqn.(\ref{64}). $\emu$ enters as the vierbein relating the two basis
systems as
$\hat{e}_{\al}=\emu\pa_{\mu}$ and $c_{\al\be}\,^{\ga}\equiv\hga$ as the
anholonomy
coefficients fulfiling
$[\hat{e}_{\al},\hat{e}_{\be}]=c_{\al\be}\,^{\ep}\hat{e}_{\ep}$.
The connection coefficients w.r.t. the frame $\hat{e}_{\al}$ are then to be
identified
as ${\it\Gamma}^{\ga}\,_{\al}\,^{\de}\equiv -\bbga$. Comparison with
eqn.(\ref{31})
shows that they transform indeed in the usual way. Note that the antisymmetry
of $\bbga$
w.r.t. $\ga$ and $\de$ translates into the metric compatibility condition for
the
connection ${\it\Gamma}$. Hence, the {\bf P\/} gauge fields are always related
to metric
connections.

Introducing next the one-form basis $\hat{\th}^{\al}$ dual to $\hat{e}_{\al}$
in
cotangential space we can turn over to the Cartan formalism defining the
connection one-form
$\om^{\ga}\,_{\de}\equiv{\it\Gamma}^{\ga}\,_{\al\de}\hat{\th}^{\al}$ and may
apply the
subsequent calculus.

The components of the Riemann tensor are defined by the second of Cartan's
structure
equations
\beq d\om^{\ga}\,_{\de}+\om^{\ga}\,_{\ep}\wedge\om^{\ep}\,_{\de}\equiv
R_{\it\Gamma}^{\ga}\,_{\de}=\frac{1}{2}R_{\it\Gamma}^{\ga}\,_{\de\al\be}\,
\hat{\th}^{\al}\wedge\hat{\th}^{\be} \eeq
and related as $R_{\it\Gamma}^{\ga\de}\,_{\al\be}=-{\tilde
R}^{\ga\de}\,_{\al\be}$ to
the field strength components introduced in eqn.(\ref{38}).
The components of the torsion tensor given by the first Cartan structure
equation
\beq d\hat{\th}^{\ga}+\om^{\ga}\,_{\ep}\wedge\hat{\th}^{\ep}\equiv
T_{\it\Gamma}^{\ga}=\frac{1}{2}T_{\it\Gamma}^{\ga}\,_{\al\be}\hat{\th}^{\al}
\wedge\hat{\th}^{\be} \eeq
are related to the components of $T$ introduced in eqn.(\ref{36}) as
$T_{\it\Gamma}^{\ga}\,_{\al\be}=-T_{\al\be}\,^{\ga}$. We remark that the
components of the tensor $K$ introduced in eqn.(49) translate into the
contortion
components $K_{\it\Gamma}^{\al}\,_{\ga\de}=-K^{\al}\,_{\ga\de}$. The gauge
potential
$C_{\al}=\frac{i}{4}\ccga\Si_{\ga\de}$ thus corresponds just to the torsion
free
Levi-Civit$\grave{\mbox{a}}$ connection
${\it\Gamma}_{L.C.}^{\ga}\,_{\al}\,^{\de}=-\ccga$.
The {\bf P\/} covariant derivative finally becomes
${\tilde\nabla}_{\al}\equiv\emu(\pa_{\mu}
-\frac{i}{4}{\it\Gamma}^{\ga}\,_{\mu}\,^{\de}\Si_{\ga\de})$,
i.e. the coordinate invariant and $SO(1,3)$ covariant derivative introduced
e.g. in
general relativity \cite{naka}.

Hence, we have established the physical equivalence of our formulation to the
geometrical
introduction of gravitational interactions in the general theory of relativity
relying
on the principle of equivalence. This allows us to interpret the fundamental
gauge
fields $\bga$ and $\bbga$ finally as gravitational potentials. The
incorporation of
the principle of equivalence in the present approach has been discussed in
\cite{wie}.

\section{Matter partition functions in gauge field backgrounds and their
scaling
behaviour}

\paragraph{}
In this section we express the scaling behaviour of the one-loop partition
functions
for the scalar, spinor and vector fields in the presence of $\ega$ and $\bbga$
in
terms of the $\ze$-function belonging to the appropriate matter fluctuation
operators.

The assumption that the interactions of the {\bf P\/} gauge fields with the
different
matter fields are renormalizable imposes strong conditions on the classical
gauge field dynamics. For
let us suppose that a given theory for a matter field and the gauge fields
$\bga$
and $\bbga$ is perturbatively renormalizable. Then we know that the change of
the
partition function of the whole system under rescaling can be absorbed in its
classical action yielding at most a nontrivial scale dependence of the
different
couplings, masses and wavefunction normalizations. Hence, the explicit
computation
of the change of the one-loop matter partition functions under rescaling will
allow us
to constrain the classical gauge field dynamics.

As technical subtleties arising in the necessary computations have already been
discussed elsewhere \cite{wip} we may turn to the evaluation of the one-loop
partition
functions and their changes under rescaling for the locally {\bf P\/} invariant
scalar, spinor and vector theories introduced in section 5.

The contribution of the scalar field to the partition function is given by
\beq {\cal Z}_{\va}[e]=\int{\cal D}\va\, e^{iS_{M}(\va;e)}. \eeq
Note that we omit possible normalizations in order to obtain the most general
renormalization structure later. After a partial integration we may rewrite
eqn.(\ref{56}) for $S_{M}$ as
\beq S_{M}(\va;e)=\frac{1}{2}\left(\va, M_{\va}(e)\va\right)_{e} \eeq
introducing the hermitean hyperbolic fluctuation operator
\beq \label{69} M_{\va}(e)\equiv-\nabla_{\al}\nabla^{\al}-m^{2}. \eeq

Performing next the Gaussian integration formally yields
\beq {\cal Z}_{\va}[e]=e^{-\frac{1}{2}\log\det M_{\va}(e)}. \eeq
As we are only interested in the behaviour of ${\cal Z}_{\va}[e]$ under
rescaling
the most suited renormalization of the ultraviolet divergent determinants above
is
based on the $\zeta-$function as it is a manifestly gauge invariant technique.

The scalar contribution to the partition function normalized at scale $\mu$
becomes with the use of eqn.(\ref{b1}) from appendix B
\beq {\cal Z}_{\va}[\mu;e]=e^{\frac{1}{2}\,\zeta'(0;\mu;M_{\va}(e))}. \eeq
Let us finally consider this contribution at the new scale ${\tilde\mu}=\la\mu$
and
determine the corresponding change of $\cal{Z}_{\va}$. With the help of
eqn.(\ref{b4})
this change becomes
\beq \label{72} {\cal Z}_{\va}[{\tilde\mu};e]={\cal Z}_{\va}[\mu;e]\cdot
e^{\log\la\cdot\zeta(0;\mu;M_{\va}(e))}. \eeq

We turn to the contribution of the spinor to the partition function. It is
given by the
Grassmann functional integral
\beq {\cal Z}_{\psi}[e,B]=\int{\cal D}\psiba{\cal D}\psi\,
e^{iS_{M}(\psiba,\psi;e,B)}. \eeq
As $S_{M}$ is already of the usual quadratic form we may perform the Grassmann
integral and formally obtain
\beq {\cal Z}_{\psi}[e,B]=e^{\frac{1}{2}\log\det M_{\psi}(e,B)}. \eeq
The hyperbolic fluctuation operator in the spinor case is obtained as usual by
squaring
the Dirac operator introduced in eqn.(\ref{59})
\beq
M_{\psi}(e,B)\equiv-\ga^{\al}({\tilde\nabla}_{\al}-\frac{1}{2}T_{\ga\al}\,^{\ga})
\cdot\ga^{\be}({\tilde\nabla}_{\be}-\frac{1}{2}T_{\de\be}\,^{\de})-m^{2} \eeq
and is hermitean w.r.t. $(\,,\,)_{e}$ due to the occurrence of $T$. We have to
recast
$M_{\psi}$ in the form of the general second order {\bf P\/} covariant operator
considered in appendix A. Using $[{\tilde\nabla}_{\al},\ga^{\be}]=0$ and
$\ga^{\al}\ga^{\be}=\eta^{\al\be}-i\Si^{\al\be}$ we obtain
\beqq M_{\psi}(e,B)&=&-({\tilde\nabla}_{\al}-\frac{1}{2}T_{\ga\al}\,^{\ga})
({\tilde\nabla}^{\al}-\frac{1}{2}T_{\de}\,^{\al\de}) \\ &+&\frac{i}{2}
\Si^{\al\be}(-T_{\al\be}\,^{\de}{\tilde\nabla}_{\de}+{\tilde R}_{\al\be}-
{\tilde\nabla}_{\al}K_{\de\be}\,^{\de})-m^{2}, \nonumber \eeqq
where the matrix in spinor space ${\tilde R}_{\al\be}$ has been defined in eqn.
(\ref{39}). Next we write
$T_{\al}\equiv\frac{i}{4}T^{\ga\de}\,_{\al}\Si_{\ga\de}$ and
absorb the first order derivative term $-2T_{\al}{\tilde\nabla}^{\al}$ in the
second
order one. Together with the use of the Jacobi identities for the
covariant derivative ${\tilde\nabla}_{\al}$ we then find the manifestly
hermitean
result
\beqq
M_{\psi}(e,B)&=&-({\tilde\nabla}_{\al}+T_{\al}-\frac{1}{2}T_{\ga\al}\,^{\ga})
({\tilde\nabla}^{\al}+T^{\al}-\frac{1}{2}T_{\de}\,^{\al\de}) \\
&+&T_{\al}\,T^{\al}
+\frac{i}{2}\Si^{\al\be}({\tilde R}_{\al\be}+{\tilde
R}^{\de}\,_{\al\de\be})-m^{2}.
\nonumber \eeqq
To obtain the form discussed in appendix A we finally introduce
\beq D_{\al}\equiv\nabla_{\al}+B_{\al}+T_{\al} \eeq
where $B_{\al}$ shall only act on spinor indices and $C_{\al}$ in
$\nabla_{\al}$ only
on vector indices. With its use we obtain the desired form
\beqq \label{79} M_{\psi}(e,B)&=&-D_{\al}\,D^{\al}+\frac{1}{2}\nabla_{\al}
T_{\ga}\,^{\al\ga}-\frac{1}{4}T_{\ga\al}\,^{\ga} T_{\de}\,^{\al\de} \\
&+&T_{\al}\,
T^{\al}+\frac{i}{2}\Si^{\al\be}({\tilde R}_{\al\be}+{\tilde
R}^{\de}\,_{\al\de\be})
-m^{2}. \nonumber \eeqq
Choosing $B=C$ in eqn.(\ref{79}) finally reduces $M_{\psi}$ to the much simpler
form
\beq M_{\psi}(e,C)=-D_{\al}\,D^{\al}
+\frac{i}{2}\Si^{\al\be}R_{\al\be}-m^{2}. \eeq

We now use eqn.(\ref{b1}) from appendix B to give the spinor contribution to
the
partition function normalized at scale $\mu$
\beq {\cal Z}_{\psi}[\mu;e,B]=e^{-\frac{1}{2}\,\zeta'(0;\mu;M_{\psi}(e,B))}.
\eeq
With the help of eqn.(\ref{b4}) we may finally express the change of
$\cal{Z}_{\psi}$
corresponding to a change of scale ${\tilde\mu}=\la\mu$ as
\beq \label{82} {\cal Z}_{\psi}[{\tilde\mu};e,B]={\cal Z}_{\psi}[\mu;e,B]\cdot
e^{-\log\la\cdot\zeta(0;\mu;M_{\psi}(e,B))}. \eeq

The contribution of the vector field to the partition function is formally
given by
\beq \label{83} {\cal Z}_{A}[e,B]=\int{\cal D}A_{\al}\,e^{iS_{M}(A;e,B)}. \eeq
First we have to recast $S_{M}$ and obtain after partially integrating
${\tilde\nabla}_{\al}$ and re-arranging the different terms
\beqq S_{M}(A;e,B)&=&\frac{1}{2}\intf\dem A_{\ep}\,\{({\tilde\nabla}_{\al}
-T_{\ga\al}\,^{\ga}){\tilde\nabla}^{\al}\eta^{\ep\ze} \nonumber \\
&-&T^{\ep\ze}\,_{\al}{\tilde\nabla}^{\al}
-{\tilde\nabla}^{\ep}T_{\ga}\,^{\ze\ga}+{\tilde R}^{\ep\ze}+m^{2}
\eta^{\ep\ze}\}\,A_{\ze} \\
&+&\frac{1}{2}\intf\dem \nabla_{\al}A^{\al}\cdot\nabla_{\be}A^{\be} \nonumber
\eeqq
where ${\tilde R}^{\ep\ze}$ is a matrix in vector field space defined in
eqn.(\ref{39}). Using again $T_{\al}=\frac{i}{4}T^{\ep\ze}\,_{\al}\Si_{\ep\ze}$
and absorbing the first order in the second order derivative term we can write
the final result as
\beqq \label{85} S_{M}(A;e,B)&=&\frac{1}{2}(A,M_{AA}(e,B)A)_{e} \\
&+&\frac{1}{2}\intf\dem \nabla_{\al}A^{\al}\cdot\nabla_{\be}A^{\be}. \nonumber
\eeqq
In terms of the operator
\beq (D_{\al})^{\ep}\,_{\eta}\equiv\nabla_{\al}\eta^{\ep}\,_{\eta}
+(B_{\al}+\frac{1}{2}T_{\al})^{\ep}\,_{\eta} \eeq
the symmetric gauge field fluctuation operator
\beqq \label{87} M_{AA}(e,B)^{\ep\ze}&\equiv&-(D_{\al})^{\ep}\,_{\eta}
(D^{\al})^{\eta\ze}+\frac{1}{4}(T_{\al})^{\ep}\,_{\eta}(T^{\al})^{\eta\ze} \\
&+&\frac{1}{2}({\tilde\nabla}^{\ep}T_{\ga}\,^{\ze\ga}-{\tilde R}^{\ep\ze}
+{\tilde\nabla}^{\ze}T_{\ga}\,^{\ep\ga}-{\tilde R}^{\ze\ep})-m^{2}\eta^{\ep\ze}
\nonumber \eeqq
is of the general form considered in appendix A. Note that $B,T,{\tilde R}$
only act
on the vector indices $\ep,\ze,\eta$ related to inner field space and $C_{\al}$
in
$\nabla_{\al}$ only on the vector indices in derivatives.

We turn to evaluate the functional integral for ${\cal Z}_{A}[e,B]$. As we are
dealing
with a massive vector field coupled to a general gauge field $\bbga$ the action
(\ref{61}) is no longer gauge invariant under the transformations defined at
the end
of section 5 and in principle we would not have to fix a gauge. As we are
also interested in the two limiting cases where the mass vanishes and where
$B=C$
we nevertheless apply the Faddeev-Popov procedure in order to be safe in taking
the
aforementioned limits restoring $U(1)$ gauge invariance. Hence we choose a
gauge
condition $F[A^{\La}_{\al}]=G(x)$ and insert the identity
$1=\int{\cal D}\La\,\de(F[A^{\La}_{\al}]-G(x))\det M_{F}(A)$
into eqn.(\ref{83}) which becomes
\beq \label{89} {\cal Z}_{A}[e,B]=\int{\cal D}\La{\cal D}A_{\al}\,
\de(F[A^{\La}_{\al}]-G(x))\det M_{F}(A)\cdot e^{iS_{M}(A;e,B)}. \eeq
As the gauge field measure and the Faddeev-Popov determinant are gauge
invariant we
may change therein the coordinates from $A_{\al}$ to $A^{\La}_{\al}=A_{\al}
+\nabla_{\al}\La$ without affecting the result. If we express the action in the
new fields we obtain
\beqq S_{M}(A^{\La},\La;e,B)&=&\intf\dem\{-\frac{1}{4}F^{\La}_{\al\be}
F^{\La\al\be}+\frac{1}{2}m^{2}A^{\La}_{\al}A^{\La\al} \nonumber \\
&-&m^{2}A^{\La}_{\al}\cdot \nabla^{\al}\La
+\frac{1}{2}F^{\La}_{\al\be}\cdot(T^{\ep})^{\al\be}\nabla_{\ep}\La  \\
&+&\frac{1}{2}m^{2}\nabla_{\al}\La\cdot\nabla^{\al}\La
-\frac{1}{4}(T^{\ep})_{\al\be}\,\nabla_{\ep}\La\cdot(T^{\eta})^{\al\be}\,
\nabla_{\eta}\La\} \nonumber \eeqq
displaying the $T$- and the $\La$-dependence explicitly. We may now change the
variable $A^{\La}_{\al}\ar A_{\al}$ everywhere in the functional integral
(\ref{89}).
To rewrite it in Gaussian form we recast the result as a quadratic form in
$\Ph\equiv(A_{\al},\La)$-space obtaining
\beqq S_{M}(\Ph;e,B)&=&\frac{1}{2}(\Ph,M_{\Ph}(e,B)\Ph)_{e} \\
&+&\frac{1}{2}\intf\dem \nabla_{\al}A^{\al}\cdot\nabla_{\be}A^{\be}. \nonumber
\eeqq
The fluctuation operator in $\Ph$-space has four entries
\beq M_{\Ph}(e,B)=\left(\begin{array}{cc}
M_{AA}^{\ep\ze} & M_{A\La}^{\ep} \\ M_{\La A}^{\ze} & M_{\La\La} \\
\end{array}\right) \eeq
and is hermitean w.r.t. $(\,,\,)_{e}$. $M_{AA}(e,B)^{\ep\ze}$ is given in
eqn.(\ref{87}) and the three other elements are found to be
\beqq M_{A\La}(e,B)^{\ep}&\equiv&({\tilde\nabla}_{\eta}-T_{\ze\eta}\,^{\ze})
(T^{\th})^{\eta\ep}\nabla_{\th}+m^{2}\nabla^{\ep} \\
M_{\La
A}(e,B)^{\ze}&\equiv&\nabla_{\eta}(T^{\eta})^{\th\ze}{\tilde\nabla}_{\th}
-m^{2}\nabla^{\ze} \nonumber \eeqq
and
\beq M_{\La\La}(e,B)\equiv-\frac{1}{2}\nabla_{\ep}(T^{\ep})_{\eta\th}
(T^{\ze})^{\eta\th}\nabla_{\ze}+m^{2}\nabla_{\ep}\nabla^{\ep}. \eeq
Note that these three operators are of a more general form than those
considered in
appendix A such that their corresponding heat kernel coefficients would have to
be
computed in a different way.

To obtain the final form of the gauge fixed functional we multiply with the
Gaussian
weight $e^{-\frac{i}{2}\intf\dem G^{2}}$ and integrate out the auxiliary field
$G$
with the result
\beq \label{95} {\cal Z}_{A\,gf}[e,B]=\int{\cal D}\Ph\,
\det M_{F}(e,B)\cdot e^{\frac{i}{2}(\Ph,M_{\Ph}(e,B)\Ph)_{e}}. \eeq
Here we made use of the background gauge condition
\beq F[A_{\al}]\equiv -\nabla_{\al}A^{\al} \eeq
to get rid of the last term in eqn.(\ref{85}) for $S_{M}$. The corresponding
Faddeev-Popov operator is then independent of $m,B,K$ and $A$ itself as in free
QED
\beq \label{97}M_{F}(e,B)=\frac{\de F[A^{\La}_{\al}]}{\de\La}
=-\nabla_{\al}\nabla^{\al} \eeq
and its determinant may be taken out of the functional integral (\ref{95})
being now
of Gaussian form. We can perform it and formally obtain
\beq {\cal Z}_{A\,gf}[e,B]=\det M_{F}(e,B)\cdot
e^{-\frac{1}{2}\log\det M_{\Ph}(e,B)}. \eeq
Although the $\zeta$-function technique discussed in appendix B may be used to
renormalize the $\det M_{\Ph}(e,B)$ we can not discuss the complete scaling
behaviour
of ${\cal Z}_{A\,gf}[e,B]$ as the fluctuation operator $M_{\Ph}(e,B)$ is no
longer
of the form considered in appendix A and we do not know its heat kernel
coefficient
functions.

But recasting $M_{\Ph}(e,B)$ as a product
\beq M_{\Ph}(e,B)=\left(\begin{array}{cc} M_{AA} & 0 \\ 0 & 1 \\
\end{array}\right)
\left(\begin{array}{cc} 1 & (M^{-1}_{AA})\,M_{A\La} \\ M_{\La A} & M_{\La\La}
\\
\end{array}\right) \eeq
we may split off the contribution coming from $\det M_{AA}(e,B)$ and obtain for
the
result regularized at scale $\mu$
\beq {\cal Z}_{A\,gf}[\mu;e,B]=e^{-\zeta'(0;\mu;M_{F}(e,B))
+\zeta'(0;\mu;M_{AA}(e,B))+\small\mbox{o.t.}} \eeq
where $\small\mbox{o.t.}$ denotes the other terms present due to nonvanishing
$T$
and $m$. At the new scale ${\tilde\mu}=\la\mu$ we then find
\beqq \label{101} {\cal Z}_{A\,gf}[{\tilde\mu};e,B]&=&{\cal
Z}_{A\,gf}[\mu;e,B]\cdot
e^{-2\log\la\cdot\zeta(0;\mu;M_{F}(e,B))} \nonumber \\
& &\cdot e^{\log\la\cdot\zeta(0;\mu;M_{AA}(e,B))+\small\mbox{o.t.}}. \eeqq
In particular we are now safe taking the limits $m=0$ and $B=C$ where all the
extra
terms simply drop out. The Faddeev-Popov operator (\ref{97}) does not change
whereas
the gauge field fluctuation operator displayed in eqn.(\ref{87}) reduces to the
simple form
\beq M_{AA}(e,C)^{\ep\ze}=-(D_{\al})^{\ep}\,_{\eta}
(D^{\al})^{\eta\ze}-R^{\ep\ze}. \eeq

\section{Renormalizability and the dynamics of the gauge fields. The minimal
gravitational action}

\paragraph{}
In this section we evaluate the $\zeta$-functions yielding the rescaling
changes
in terms of the {\bf P\/} gauge fields. We then determine a minimal gauge field
action compatible with renormalizability requirements.

In the previous section we expressed the changes under rescaling of the
one-loop
partition functions for scalar, spinor and vector fields in terms of different
$\zeta$-functions. Renormalizability of any theory including dynamical gauge
fields
requires now at least that these anomalous contributions, which are local
polynomials
in $\ega$ and $\bbga$ and their derivatives, may be absorbed in the classical
action
for the gauge fields $\ega$ and $\bbga$. Hence, to determine explicitly a
minimal
gauge field dynamics consistent with renormalizability we
finally have to evaluate the different $\zeta$-functions.

Let us begin with the scalar field. The corresponding fluctuation operator is
given in
eqn.(\ref{69}) and is of the form of the general operator (\ref{a1}) in
appendix A
if we choose $A_{\al}=0, E=-m^{2}$ there. The coefficient function
$U_{m}(x)\equiv
\tr_{S} c_{2}(x)$ obtained from eqn.(\ref{a22}) reduces to a quite simple form
\beqq \label{102}
U_{m}&=&-\frac{1}{30}\nabla_{\ga}\nabla^{\ga}R^{\al\be}\,_{\al\be}
+\frac{1}{72}R_{\al\be}\,^{\al\be}\cdot R_{\ga\de}\,^{\ga\de} \nonumber \\
&+&\frac{1}{180}R_{\al\be\ga\de}\cdot R^{\al\be\ga\de}
-\frac{1}{180}R_{\al\ga}\,^{\al}\,_{\de}\cdot R_{\be}\,^{\ga\be\de} \\
&-&\frac{1}{6}m^{2}\cdot R^{\al\be}\,_{\al\be}+\frac{1}{2}m^{4}. \nonumber
\eeqq
With the use of eqn.(\ref{b6}) from appendix B we next obtain the value of
$\zeta(0;\mu;M_{\va}(e))$ as the integral over $\tr_{S} c_{2}(x)$. Its
insertion into
eqn.(\ref{72}) finally yields the anomalous term in the scalar case.

Next we turn to the spinor sector. The operator (\ref{a1}) of appendix A
coincides
with the spinor fluctuation operator given in eqn.(\ref{79}) provided that we
set
\beq A_{\al}\equiv B_{\al}+T_{\al}=\frac{i}{4}\Si_{\ga\de}(B_{\al}\,^{\ga\de}
+T^{\ga\de}\,_{\al}), \eeq
where $\Si_{\ga\de}$ acts on the spinor indices only, and
\beq E\equiv V_{1}+V_{2}+V_{3}-m^{2}. \eeq
Here we introduced
\beqq V_{1}&=&-\frac{1}{8}\Si^{\al\be}\Si^{\ga\de}\,V_{1\ga\de\al\be},\quad
V_{1\ga\de\al\be}={\tilde R}_{\ga\de\al\be}
+\frac{1}{2}T_{\ga\de\eta}T_{\al\be}\,^{\eta}, \nonumber \\
V_{2}&=&\frac{i}{4}\Si^{\al\be}\,V_{2\al\be},\quad
V_{2\al\be}=\frac{1}{2}(V_{1}^{\eta}\,_{\al\eta\be}-V_{1}^{\eta}\,_{\be\eta\al}), \\
V_{3}&=&\frac{1}{2}{\tilde\nabla}_{\al}T_{\ga}\,^{\al\ga}
-\frac{1}{4}T_{\ga\al}\,^{\ga} T_{\de}\,^{\al\de}. \nonumber \eeqq
We also need the field strength $F_{\al\be}$ corresponding to $A_{\al}$ defined
in eqn.(\ref{a21})
\beqq F_{\al\be}&=&\frac{i}{4}\Si^{\ga\de}\,F_{\ga\de\al\be} \nonumber \\
F_{\ga\de\al\be}&=&{\tilde R}_{\ga\de\al\be}+{\tilde\nabla}_{\al}T_{\ga\de\be}
-{\tilde\nabla}_{\be}T_{\ga\de\al} \\
&+&T_{\al\be}\,^{\eta}T_{\ga\de\eta}+T^{\eta}\,_{\ga\al}T_{\eta\de\be}
-T^{\eta}\,_{\ga\be}T_{\eta\de\al}. \nonumber \eeqq
Now we insert the above expressions into eqn.(\ref{a22}) for $c_{2}(x)$ and
take the
Dirac trace
\beqq \label{107} \tr_{D}
c_{2}&=&4U_{m}+(\frac{1}{6}R^{\al\be}\,_{\al\be}-m^{2})
\cdot(4V_{3}-V_{1}^{\ga\de}\,_{\ga\de}) \nonumber \\
&-&\frac{1}{6}\nabla_{\al}\nabla^{\al}(4V_{3}-V_{1}^{\ga\de}\,_{\ga\de})
-\frac{1}{24}F_{\al\be\ga\de}\cdot F^{\al\be\ga\de} \\
&-&V_{3}\cdot V_{1}^{\ga\de}\,_{\ga\de}-\frac{1}{8}V_{2\al\be}\cdot
V_{2}^{\al\be}+\frac{1}{8}V_{1}^{\al\be}\,_{\al\be}\cdot
V_{1}^{\ga\de}\,_{\ga\de} \nonumber \\
&+&2V_{3}^{2}+\frac{1}{8}V_{1\al\be\ga\de}\cdot(V_{1}^{\al\be\ga\de}
+V_{1}^{\ga\de\al\be}+V_{1}^{\ga\be\de\al}). \nonumber \eeqq
With the use of eqn.(\ref{b6}) from appendix B we next obtain the value of
$\zeta(0;\mu;M_{\psi}(e,B))$
which finally yields the anomalous term in eqn.(\ref{82}) in the spinor case.
We
remark that for $T\ne 0$ this result contains a huge number of different terms
if
we recast it in the natural variables ${\tilde R}$ and $T$. Only for $T=0$ it
reduces to a simple form with
\beqq \label{108} \tr_{D} c_{2}&=&\frac{1}{30}\nabla_{\ga}\nabla^{\ga}
R^{\al\be}\,_{\al\be}
+\frac{1}{72}R_{\al\be}\,^{\al\be}\cdot R_{\ga\de}\,^{\ga\de} \nonumber \\
&-&\frac{7}{360}R_{\al\be\ga\de}\cdot R^{\al\be\ga\de}
-\frac{1}{45}R_{\al\ga}\,^{\al}\,_{\de}\cdot R_{\be}\,^{\ga\be\de} \\
&+&\frac{1}{3}m^{2}\cdot R^{\al\be}\,_{\al\be}+2m^{4}. \nonumber \eeqq

In the vector case we have to evaluate both the $\zeta$-functions belonging to
the ghost operator $M_{F}$ and the vector operator $M_{AA}$. The former has
been
obtained in eqn.(\ref{95}) and coincides with the operator (\ref{a1}) of
appendix A
if we choose $A_{\al}=E=0$ whereas the latter, given in eqn.(\ref{87}),
coincides
with the operator (\ref{a1}) provided that we set
\beq A_{\al}\equiv B_{\al}+\frac{1}{2}T_{\al}=\frac{i}{4}\Si_{\ga\de}
(B_{\al}\,^{\ga\de}+\frac{1}{2}T^{\ga\de}\,_{\al}), \eeq
where $\Si_{\ga\de}$ acts on inner vector indices only, and
\beq E^{\ep\ze}\equiv V^{\ep\ze}-m^{2}\eta^{\ep\ze}. \eeq
Here we set
\beq V^{\ep\ze}=\frac{1}{2}({\tilde\nabla}^{\ep}T_{\ga}\,^{\ze\ga}
-{\tilde R}_{\ga}\,^{\ze\ga\ep}+{\tilde\nabla}^{\ze}T_{\ga}\,^{\ep\ga}
-{\tilde R}_{\ga}\,^{\ep\ga\ze})-\frac{1}{4}T^{\ep}\,_{\ga\de}\,T^{\ze\ga\de}.
\eeq
The field strength $F_{\al\be}$ corresponding to $A_{\al}$ is found to be
\beqq F_{\al\be}&=&\frac{i}{4}\Si^{\ga\de}\,F_{\ga\de\al\be} \nonumber \\
F_{\ga\de\al\be}&=&{\tilde R}_{\ga\de\al\be}+\frac{1}{2}{\tilde\nabla}_{\al}
T_{\ga\de\be}-\frac{1}{2}{\tilde\nabla}_{\be}T_{\ga\de\al} \\
&+&\frac{1}{2}T_{\al\be}\,^{\eta}T_{\ga\de\eta}+\frac{1}{4}T^{\eta}\,_{\ga\al}
T_{\eta\de\be}-\frac{1}{4}T^{\eta}\,_{\ga\be}T_{\eta\de\al}. \nonumber \eeqq
Inserting the above expressions into eqn.(\ref{a22}) for $c_{2}(x)$ and taking
the
respective traces we get in the ghost case the same result as in the scalar one
for $m=0$
\beq \label{113} \tr_{G} c_{2}=U_{0}, \eeq
whereas the result in the vector case is
\beqq \label{114} \tr_{M}
c_{2}&=&4U_{m}+(\frac{1}{6}R^{\al\be}\,_{\al\be}-m^{2})
\cdot V_{\ga}\,^{\ga} \nonumber \\
&-&\frac{1}{6}\nabla_{\al}\nabla^{\al}V_{\ga}\,^{\ga}
-\frac{1}{12}F_{\al\be\ga\de}\cdot F^{\al\be\ga\de} \\
&+&\frac{1}{2}V^{\al\be}\cdot V_{\al\be}. \nonumber \eeqq
With the use of eqn.(\ref{b6}) from appendix B we next obtain the values of
$\zeta(0;\mu;M_{F}(e))$ and $\zeta(0;\mu;M_{AA}(e,B))$
which finally yield the anomalous terms in eqn.(\ref{101}) for the vector case.
Again, for $T\ne 0$ the result (\ref{114}) contains a huge number of different
terms if
we recast it in the natural variables ${\tilde R}$ and $T$. Only in the $U(1)$
gauge
invariant case, for $T=m=0$, it reduces to the simple form
\beqq \label{115} \tr_{M} c_{2}&=&\frac{1}{30}\nabla_{\ga}\nabla^{\ga}
R^{\al\be}\,_{\al\be}-\frac{1}{9}R_{\al\be}\,^{\al\be}\cdot
R_{\ga\de}\,^{\ga\de} \\
&-&\frac{11}{180}R_{\al\be\ga\de}\cdot R^{\al\be\ga\de}
-\frac{43}{90}R_{\al\ga}\,^{\al}\,_{\de}\cdot R_{\be}\,^{\ga\be\de}. \nonumber
\eeqq
The results eqns.(\ref{102}), (\ref{108}) and (\ref{115}) for $T=0$ are
contained
in \cite{chri} as special cases.

In eqns.(\ref{102}), (\ref{107}), (\ref{113}) and (\ref{114}) we have
explicitly obtained
the different anomalous contributions to the rescaled partition functions as
local
{\bf P\/} gauge invariant polynomials in the fields $\ega$ and $\bbga$. As
discussed
above, they also must be present in any classical gauge field dynamics
consistent with
renormalizability of the matter sectors. Hence, we are finally led to construct
a minimal action for the gauge fields just in terms of these {\bf P\/} gauge
invariant
polynomials. Note that this reasoning yields in the case of non-abelian matrix
groups
indeed the usual Yang-Mills action.

For $T\ne 0$ we restrict ourselves to the contributions of $O(\pa^{0},\pa^{2})$
in the derivatives and obtain as minimal classical action to this order
\beqq \label{116} S_{G}(e,B)&=&\int\dem\{\Lambda-\frac{1}{\kappa^{2}}\cdot
{\tilde R}_{\al\be}\,^{\al\be}+\be_{1}\cdot T_{\ga\al}\,^{\ga}
T_{\de}\,^{\al\de} \\
&+&\be_{2}\cdot T_{\al\be\ga}T^{\al\be\ga}+\be_{3}\cdot T_{\al\be\ga}
T^{\ga\al\be}
+O(\pa^{4})\}, \nonumber \eeqq
skipping possible total divergence terms. Here we have to introduce different
couplings $\kappa,\be_{1},\be_{2},\be_{3}$ and the constant $\Lambda$ which are
independently renormalized by the one-loop contributions we determined above.
Note
that our reasoning automatically enforces a cosmological constant as to be
expected
from general renormalization considerations. The action eqn.(\ref{116})
describes the
classical gauge field dynamics correctly at sufficiently low momentum scales
and small
values of the couplings. Nevertheless, only a dynamics containing the huge
number of
different $O(\pa^{4})$ terms as well, coming along with the same number of
independent
couplings, will be consistent with renormalizability \cite{buc}.

If we set $T=0$ the minimal classical action must contain the terms
\beqq \label{117} S_{G}(e)&=&\int\dem\{\Lambda-\frac{1}{\kappa^{2}}\cdot
R_{\al\be}\,^{\al\be}+\al_{1}\cdot R_{\al\be}\,^{\al\be}\cdot
R_{\ga\de}\,^{\ga\de} \\
&+&\al_{2}\cdot R_{\al\ga}\,^{\al}\,_{\de}\cdot R_{\be}\,^{\ga\be\de}
+\al_{3}\cdot R_{\al\be\ga\de}\cdot R^{\al\be\ga\de}, \nonumber \eeqq
if discarding total divergencies. The couplings
$\kappa,\al_{1},\al_{2},\al_{3}$
and the constant $\Lambda$ obtain again contributions from the one-loop scale
anomalies which have been determined above. We emphasize that $S_{G}$ is an
action
for gauge fields defined on the Minkowski spacetime ({\bf R\/}$^{4}$,$\eta$)
and
is invariant on one hand under local {\bf P\/} gauge transformations, on the
other
hand under global Poincar$\acute{\mbox{e}}$ transformations reflecting the
symmetries
of the underlying spacetime.

Important aspects of the quantized theory (\ref{117}) such as one-loop
divergencies
and $\beta$-functions and its unitarity problems are discussed in \cite{buc}
and
references given there.

\section{Conclusions}

\paragraph{}
Based on the complementary conception of Poincar$\acute{\mbox{e}}$ symmetry as
a
purely inner symmetry we have developed a {\bf P\/} gauge theory of
gravitation.
The gravitational interaction is mediated by gauge fields defined on a fixed
Minkowski
spacetime. Their dynamics has been determined imposing consistency requirements
with
renormalization properties of matter fields in gravitational backgrounds. In an
appropriate low energy limit it reduces to a form yielding the same
observational
predictions as made in general relativity.

In our conception there is no direct interrelation between gravity and the
structure
of spacetime. E.g., only if asking about the behaviour of rods and clocks at
the
classical level one is led to introduce an effective metric containing the
desired
information \cite{wie}. On the other hand, at the quantum level it may
conceptually be easier to
deal with a field theoretical description of gravitation free of any
geometrical
aspects.

This may shed some new light on questions related to the causality structure of
spacetime at the quantum level, or the question of energy-momentum carried by
the
gravitational fields. Namely, the separation of the local gauge group {\bf P\/}
from
the global Poincar$\acute{\mbox{e}}$ symmetry group of the underlying Minkowski
spacetime will allow us to obtain the energy-momentum carried by the gauge
potentials
in the usual Noether way.

In the determination of the scaling behaviour of the one-loop vector field
partition
function we obtained fluctuation operators of a more general form than usually
investigated. Working out the coefficient functions occuring in the asymptotic
expansion of the corresponding heat kernels poses an interesting technical
problem
in its own and is a necessary ingredient of a determination of the full scaling
behaviour for $T\ne 0, m\ne 0$.

The most serious drawback of the present approach is of course the necessity of
including the terms quadratic in the field strength in the classical gauge
field
action. Although the corresponding quantum theory is known to be
renormalizable,
the occurrence of negative energy or negative norm ghost states has destroyed
up
to now any attempt of establishing unitarity and hence a physical
interpretation of
the theory \cite{buc}.

\section*{Acknowledgments}

\paragraph{}
This work has partially been supported by Schweizerischer Nationalfonds. I am
indebted
to S. D\"urr and R. B\"artschi for their encouragement, to F. Krahe for
intensive
discussions and a careful reading of the manuscript. L. O'Raifeartaigh helped
me
clarifying some group theoretical aspects. Helpful remarks of J. Kambor, J.
Stern and
A. Wipf are acknowledged.

\appendix

\section{Heat kernel coefficients of {\bf P\/} gauge covariant differential
operators}

\paragraph{}
In this appendix we determine the heat kernel coefficients $c_{1}$ and $c_{2}$
belonging to a general hermitean {\bf P\/} covariant second order differential
operator $M$ defined on the $d$-dimensional Minkowski spacetime
({\bf R\/}$^{d}$,$\eta$). We adapt here well-known techniques developed in a
geometrical context to our case \cite{dewi} - \cite{brag}.

Let us consider the {\bf P\/} covariant hermitean operator
\beq \label{a1} M=-D_{\al}D^{\al}+E,\quad\quad\quad
D_{\al}=\nabla_{\al}+A_{\al}. \eeq
$\nabla_{\al}=d_{\al}+C_{\al}$ is the {\bf P\/} covariant derivative defined in
eqn.(\ref{42}). We emphasize that
$C_{\al}=\frac{i}{4}C_{\al}\,^{\ga\de}\Si_{\ga\de}$
is throughout understood to be adjusted to the Lorentz group representation it
acts
upon to ensure the covariant transformation properties of $D_{\al}$. The
anti-hermitean matrix-valued four-vector $A_{\al}$, on the other hand, is kept
fixed.
Finally, $E$ is a general hermitean matrix field.

The heat kernel $K(is;x,y),\,s>0$, belonging to $M_{x}$ fulfils
\beq \label{a6} \left( \frac{\pa}{\pa(is)}+M_{x}\right) K(is;x,y)=0 \eeq
together with the initial condition $\lim_{s\to
0}K(is;x,y)=\frac{1}{\dem}\,\de(x-y)$.
We are interested in the small $s$-expansion of $K(is;x,y)$ in the coincidence
limit
$y\to x$. Asymptotically this expansion is of the form
\beq \label{a7} K(is;x,y)\,\,\,{\sim\!\!\!\!\!\!\!\!^{s\to 0}}
\frac{i}{(4\pi is)^{\frac{d}{2}}}\,
e^{-\frac{r^{2}(x,y)}{4is}}\,\sum_{k=0}^{\infty}(is)^{k}c_{k}(x,y). \eeq
Hence, the task is to evaluate $r^{2}(x,y)$ and the coefficient functions
$c_{k}(x,y)$. Inserting the expansion (\ref{a7}) in eqn.(\ref{a6}) and equating
equal
powers of $s$ we obtain the three {\bf P\/} covariant relations
\beqq \label{a8} & &r^{2}(x,y)=\frac{1}{4}\nabla_{\al}r^{2}(x,y)
\cdot\nabla^{\al}r^{2}(x,y), \\
& & \label{a9} \frac{1}{2}\nabla_{\al}r^{2}(x,y)\cdot D^{\al}c_{0}(x,y)
=\left\{ \frac{d}{2}-\frac{1}{4}\nabla_{\al}\nabla^{\al}r^{2}(x,y)\right\}
\cdot c_{0}(x,y), \\
& & \label{a10} \frac{1}{2}\nabla_{\al}r^{2}(x,y)\cdot D^{\al}c_{k+1}(x,y) \\
& &\quad\quad=\left\{
\frac{d}{2}-k-1-\frac{1}{4}\nabla_{\al}\nabla^{\al}r^{2}(x,y)
\right\}\cdot c_{k+1}(x,y)-M_{x} c_{k}(x,y). \nonumber \eeqq
The first equation allows to evaluate $r^{2}$ and all its covariant derivatives
at
$y=x$ whereas the two other relations (\ref{a9}) and (\ref{a10}) allow a
recursive
determination of $c_{k}$ and all its covariant derivatives again at $y=x$.
$c_{0}=1$
at $y=x$ ensures the correct initial condition. Note the introduction of the
shorthand notation $f(x)\equiv f(x,x)$ for functions taken in the coincidence
limit
$y=x$.

We turn to the calculation of $r^{2}$ and its covariant derivatives.
Differentiation of eqn.(\ref{a8}) leads to the relations
\beqq \label{a11}
2\nabla_{\al}r^{2}&=&\nabla_{\al\eta}r^{2}\cdot\nabla^{\eta}r^{2}, \\
\label{a12}
2\nabla_{\be\al}r^{2}&=&\nabla_{\be\al\eta}r^{2}\cdot\nabla^{\eta}r^{2}
+\nabla_{\al\eta}r^{2}\cdot\nabla_{\be}\,^{\eta}r^{2}, \\
\label{a13}
2\nabla_{\ga\be\al}r^{2}&=&\nabla_{\ga\be\al\eta}r^{2}\cdot\nabla^{\eta}r^{2}
+\nabla_{\be\al\eta}r^{2}\cdot\nabla_{\ga}\,^{\eta}r^{2} \\
&+&\nabla_{\ga\al\eta}r^{2}\cdot\nabla_{\be}\,^{\eta}r^{2}
+\nabla_{\al\eta}r^{2}\cdot\nabla_{\ga\be}\,^{\eta}r^{2}.
\nonumber \eeqq
Here we introduced the shorthand notations
$\nabla_{\be\al}=\nabla_{\be}\nabla_{\al}$
etc.. As the initial value is $r^{2}(x)=0$, the relation (\ref{a11}) leads to
\beq \label{a14} \nabla_{\al}r^{2}(x)=0 \eeq
which is consistent with (\ref{a8}). The use of eqn.(\ref{a14}) in the second
relation
(\ref{a12}) yields now
$2\nabla_{\be\al}r^{2}(x)=\nabla_{\al\eta}r^{2}(x)\cdot\nabla_{\be}\,^{\eta}r^{2}(x)$
and is solved by
\beq \nabla_{\be\al}r^{2}(x)=2\eta_{\be\al}.\eeq
As $\eta_{\be\al}$ is the only second rank tensor with the desired covariance
properties this solution is in fact unique. Using the above results the third
relation (\ref{a13}) becomes
$\nabla_{\ga\be\al}r^{2}(x)=\nabla_{\be\al\ga}r^{2}(x)+\nabla_{\ga\al\be}r^{2}(x)
+\nabla_{\ga\be\al}r^{2}(x)$. To solve it we commute the covariant derivatives
according to eqn.(\ref{43}). This yields e.g.
$\nabla_{\ga\al\be}r^{2}(x)=\nabla_{\ga\be\al}r^{2}(x)
+\nabla_{\ga}(R_{\al\be}\,r^{2}(x))=\nabla_{\ga\be\al}r^{2}(x)$,
for $r^{2}(x)$ is a scalar and thus $R_{\al\be}\,r^{2}(x)=0$. Hence, we find
\beq \label{a16} \nabla_{\ga\be\al}r^{2}(x)=0 \eeq
expressing simply the fact that no homogeneously transforming third rank tensor
built
from $\ega$ exists. In the same way we obtain
\beq
\nabla_{\de\ga\be\al}r^{2}(x)=\frac{2}{3}(R_{\al\de\be\ga}+R_{\al\ga\be\de})
\eeq
and higher derivatives.

We turn to the computation of $c_{1}(x)$. Appropriate differentiation of the
relation
(\ref{a10}) for $k=0$ and the use of the results eqns.(\ref{a14})
-(\ref{a16}) for the covariant derivatives of $r^{2}$ lead in the limit $y\to
x$ to
\beq c_{1}(x)=D_{\al}\,^{\al}c_{0}(x)-E\cdot c_{0}(x). \eeq
Note the introduction of the shorthand notations $D_{\be\al}=D_{\be}D_{\al}$
etc..
There remain the different higher derivatives of $c_{0}$ to be determined. We
now
differentiate the relation (\ref{a9}) for $c_{0}$ and obtain together with the
results eqns.(\ref{a14})-(\ref{a16}) in the coincidence limit
\beqq D_{\al} c_{0}(x)&=&0, \\
D_{\al}\,^{\al}c_{0}(x)&=&-\frac{1}{8}\nabla_{\al}\,^{\al}\,_{\be}\,^{\be}r^{2}(x)
\cdot c_{0}(x). \nonumber \eeqq
Inserting the initial condition $c_{0}(x)=1$ finally yields
\beq \label{a20} c_{1}(x)=-\frac{1}{6}R^{\al\be}\,_{\al\be}-E. \eeq

The calculation of $c_{2}(x)$ is algebraically more involved. We thus restrict
ourselves to note that it requires the use of the commutation relation for
the covariant derivative $D_{\al}$
\beqq \label{a21} [D_{\al},D_{\be}]
&=&[\nabla_{\al},\nabla_{\be}]+\nabla_{\al}A_{\be}-\nabla_{\be}A_{\al}
+[A_{\al},A_{\be}] \\
&=&R_{\al\be}+F_{\al\be} \nonumber \eeqq
which defines the field strength $F_{\al\be}$ belonging to the gauge field
$A_{\al}$.
The Jacobi identities for the covariant derivative $\nabla_{\al}$ lead to
$R_{\eta\al\be\ga}+R_{\eta\ga\al\be}+R_{\eta\be\ga\al}=0$ and
$\nabla_{\ga}R^{\al\be}\,_{\al\be}=2\nabla_{\al}R^{\al\be}\,_{\ga\be}$
allowing then to bring the result for $c_{2}(x)$ into the simple form
\beqq \label{a22}
c_{2}(x)&=&-\frac{1}{30}\nabla_{\ga}\,^{\ga}R^{\al\be}\,_{\al\be}
+\frac{1}{72}R_{\al\be}\,^{\al\be}\cdot R_{\ga\de}\,^{\ga\de} \nonumber \\
&+&\frac{1}{180}R_{\al\be\ga\de}\cdot R^{\al\be\ga\de}
-\frac{1}{180}R_{\al\ga}\,^{\al}\,_{\de}\cdot R_{\be}\,^{\ga\be\de} \\
&+&\frac{1}{12}F_{\al\be}\cdot F^{\al\be}
+\frac{1}{6}R^{\al\be}\,_{\al\be}\cdot E \nonumber \\
&-&\frac{1}{6}[D_{\al},[D^{\al},E]]+\frac{1}{2}E^{2}. \nonumber \eeqq

We finally remark that unfortunately only $c_{1}$ has been computed directly
for an
operator $M$ with $D_{\al}={\tilde\nabla}_{\al}+A_{\al}$, where
${\tilde\nabla}_{\al}=d_{\al}+B_{\al}$ includes torsion \cite{gusy}.

\section{$\zeta-$function regularization of functional determinants and
$\zeta(0)$.}

\paragraph{}
In this appendix we define the functional determinant belonging to $M$ in terms
of
the $\zeta-$function regularization technique.  We then determine its change
under a
rescaling using the heat kernel expansion obtained in appendix A.

One can define the functional determinant of the operator $M=-D_{\al}D^{\al}+E$
introduced in appendix A to be \cite{dow}, \cite{haw}
\beq \label{b1} \log\det M\equiv -\lim_{u\to 0}\frac{d}{du}\,\zeta(u;\mu;M)
\eeq
where the generalized $\zeta-$function belonging to $M$ is given by
\beq \zeta(u;\mu;M)\equiv\mu^{2u}\Tr M^{-u}. \eeq
The scale $\mu$ at which parameters such as couplings, masses and wavefunction
normalizations have to be adjusted is introduced in order to keep the
determinant
dimensionless.

The above definition of $\zeta$ does not allow to take the $u$-derivative at
$u=0$
since the trace is defined only for Re$\,u>\frac{d}{2}$. The necessary analytic
continuation is achieved by recasting $\zeta$ as the Mellin transformation of
the
heat kernel
\beq \label{b3} \zeta(u;\mu;M)=\frac{i\mu^{2u}}{\it\Gamma(u)}\int
\limits_{0}^{\infty} ds
\,(is)^{u-1}\Tr e^{-isM} \eeq
and yields indeed the desired ultraviolet regularization.

Let us next consider the behaviour of the functional determinant under a change
of
scale ${\tilde\mu}=\la\mu$. One obtains
\beq \label{b4} \zeta'(0;{\tilde\mu};M)=\zeta'(0;\mu;M)
+2\log\la\cdot\zeta(0;\mu;M). \eeq
The change of the functional determinant under a rescaling is thus fully
determined
by $\zeta(0;\mu;M)$.

To evaluate $\zeta(0;\mu;M)$ we use the representation eqn.(\ref{b3}). It is
the
singular part of the $s$-integration in eqn.(\ref{b3}) which yields a
nonvanishing value for $\zeta(0;\mu;M)$. As this singular part comes from the
small
$s$-region we may use the expansion for the trace of the heat kernel following
from eqn.(\ref{a7}) in appendix A
\beq \Tr e^{-isM}\quad{\sim\!\!\!\!\!\!\!\!^{s\to 0}}\frac{i}{(4\pi
is)^{\frac{d}{2}}}\,
\sum_{k=0}^{\infty}(is)^{k}\int d^{d}x\dem\tr c_{k}(x). \eeq
Performing the $s$-integration in (\ref{b3}) singles out the contribution for
$k=\frac{d}{2}$ from the infinite sum and one obtains
\beq \label{b6} \zeta(0;\mu;M)=\frac{i}{(4\pi)^{\frac{d}{2}}}\,\int d^{d}x\dem
\tr c_{\frac{d}{2}}(x). \eeq

\end{document}